\documentclass[letterpaper, 10 pt, conference]{ieeeconf}
\IEEEoverridecommandlockouts
\usepackage{cite}
\usepackage{flushend}
\usepackage{amsmath,amssymb,amsfonts}
\usepackage{algorithmic}
\usepackage{graphicx}
\usepackage{textcomp}
\usepackage{xcolor}
\usepackage{multirow}
\usepackage{booktabs}

\def\BibTeX{{\rm B\kern-.05em{\sc i\kern-.025em b}\kern-.08em
    T\kern-.1667em\lower.7ex\hbox{E}\kern-.125emX}}
\begin{document}

\title{Composite Biomarker Image for Advanced Visualization in Histopathology}

\author{Abubakr Shafique$^{1,2}$, Morteza Babaie$^{1,3}$, Ricardo Gonzalez$^{4}$, \\Adrian Batten$^{5}$, Soma Sikdar$^{5}$, and H.R. Tizhoosh$^{1,2,3}$
\thanks{$^{1}$Abubakr Shafique, Morteza Babaie, and H.R. Tizhoosh are with Kimia Lab, University of Waterloo, Waterloo, ON, Canada (email:  abubakr.shafique@uwaterloo.ca)}
\thanks{$^{2}$Abubakr Shafique and H.R. Tizhoosh are also affiliated with Rhazes Lab, Department of Artificial Intelligence \& Informatics, Mayo Clinic, Rochester, MN, USA.}
\thanks{$^{3}$Hamid Tizhoosh and Morteza  Babaie are also affiliated with Vector Institute, MaRS Centre, Toronto, Canada.}
\thanks{$^{4}$Ricardo Gonzalez is affiliated with Laboratory Medicine and Pathology, Mayo Clinic, Rochester, MN, USA.}
\thanks{$^{5}$Adrian Batten and Soma Sikdar are with Department of Pathology, Grand River Hospital, Kitchener, ON, Canada.}
}
\maketitle

\begin{abstract}
Immunohistochemistry (IHC) biomarkers are essential tools for reliable cancer diagnosis and subtyping. It requires cross-staining comparison among Whole Slide Images (WSIs) of IHCs and hematoxylin and eosin (H\&E) slides. Currently, pathologists examine the visually co-localized areas across IHC and H\&E glass slides for a final diagnosis, which is a tedious and challenging task. Moreover, visually inspecting different IHC slides back and forth to analyze local co-expressions is inherently subjective and prone to error, even when carried out by experienced pathologists. Relying on digital pathology, we propose ``Composite Biomarker Image'' (CBI) in this work. CBI is a single image that can be composed using different filtered IHC biomarker images for better visualization.
We present a CBI image produced in two steps by the proposed solution for better visualization and hence more efficient clinical workflow. In the first step, IHC biomarker images are aligned with the H\&E images using one coordinate system and orientation. In the second step, the positive or negative IHC regions from each biomarker image (based on the pathologists’ recommendation) are filtered and combined into one image using a \emph{fuzzy inference system}. For evaluation, the resulting CBI images, from the proposed system, were evaluated qualitatively by the expert pathologists. The CBI concept helps the pathologists to identify the suspected target tissues more easily, which could be further assessed by examining the actual WSIs at the same suspected regions.
\end{abstract}

\section{Introduction}
\label{sec:intro}
Wide adoption of whole slide image (WSI) analysis by the pathologists community helps them to improve and speedup their diagnostic process~\cite{HAMILTON2014}. Immunohistochemistry (IHC) analysis is a pertinent tool for the detection of specific antigens in tissue sections~\cite{Duraiyan2012}. It is one of the significant technique for determining the tissue distribution of an antigen of interest in health and disease using monoclonal and polyclonal antibodies~\cite{Duraiyan2012}. In pathology, IHC is particularly essential in the subspecialties of oncologic pathology, neuropathology, and hematopathology. While basic histologic examination of tissue is considered a helpful and important component of autopsy pathology, IHC may provide more insight~\cite{roulson2005}. Many literature have discussed about the diagnostic utility of IHC in surgical pathology~\cite{leong1987}. IHC's uses have exploded in recent years as more and more molecules implicated in disease development, diagnosis, and treatment have been found~\cite{kim2016}. Combined analysis of multiple antibody expressions and tissue morphology is important for diagnosis, treatment planning and drug discovery~\cite{Abubakr2021}. Cellular proteins, morphology, and IHC biomarkers are critical tools for determining if a malignancy is benign or malignant, determining the stage and grade of a tumour, and identifying the cell type and origin of a metastasis in order to locate the primary tumour~\cite{Feldman2007}. 

\begin{figure*}[ht]
\centerline{\includegraphics[width =\textwidth]{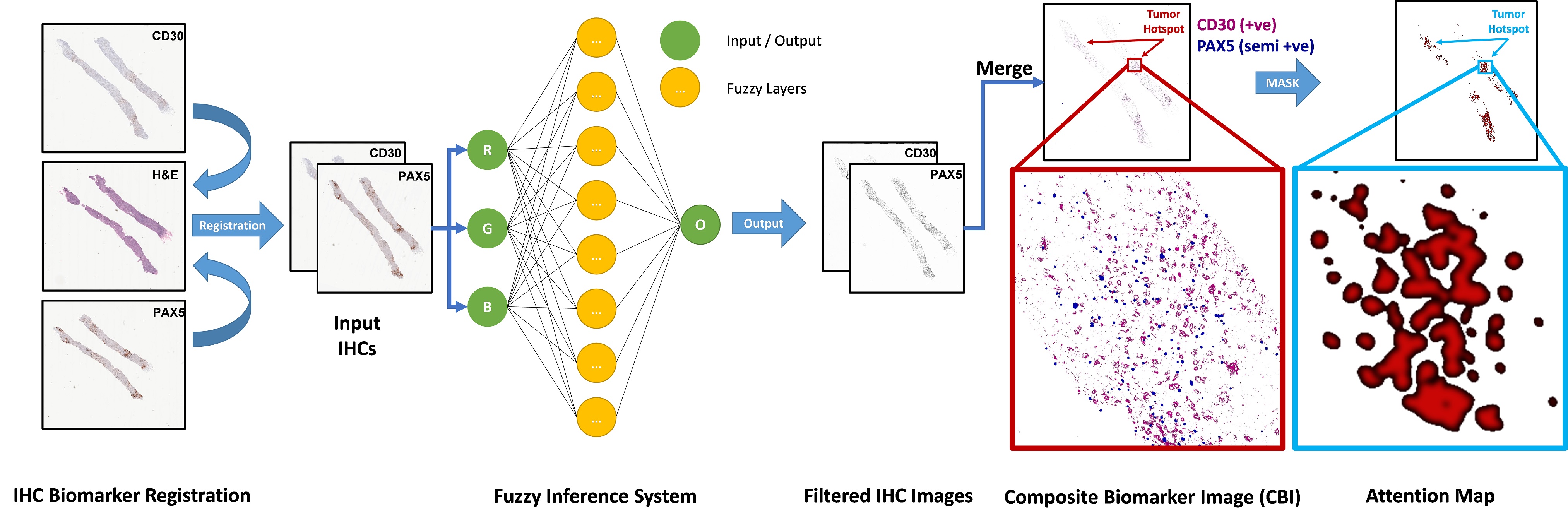}}
\caption{Graphical representation of the two step procedure in the proposed system. In the first step, PAX5, and CD30 source biomarker images are shown to be registered according to the target H\&E image. Second step shows the fuzzy inference systems used to filter the IHC biomarker images. Finally, these filtered images are fused and overlayed together to produce a final CBI image with distinct coloring, which eventually converted into an attention map.}
\label{Fig:Methods}
\end{figure*}

More than 8500 people are diagnosed with Hodgkin Lymphoma (HL) every year in United States, which is nearly 10.2\% of all lymphoma cases~\cite{Ansell2018}. Classical Hodgkin Lymphoma (CHL) represents the most common lymphoma sub-type in children and young adults~\cite{AnjaMottok2018}. CHL is prominent among lymphoid malignancies because of its histomorphological appearance~\cite{AnjaMottok2018}. It is important to identify the malignant Reed‐Sternberg cell, which is of follicular center B‐cell origin~\cite{MARAFIOTI2000}. Previously, the diagnosis of HL was based on the morphologic identification of Reed-Sternberg cells and variants in the appropriate cellular milieu~\cite{Denise2014}. Now, the diagnosis of HL relies on the joint analysis of hematoxylin and eosin (H\&E) and IHC biomarkers~\cite{Denise2014}. IHC biomarkers help pathologists to indicate the presence and expression of a chosen protein, which helps them for accurate diagnosis~\cite{Trahearn2017}. In the case of CHL, it could be identified when there is the presence of CD30 (+) Hodgkin and Reed-Sternberg cells with light expression of PAX5 (B-lymphocytes)~\cite{ValiBetts2017}. IHC biomarkers also help the pathologists to distinguish confusing cases such as  T-cell histiocyte-rich large B-cell lymphoma, peripheral T-cell lymphomas with CD30-positive cells, and reactive cases with CD30-positive immunoblasts~\cite{Carbone1992, Torlakovic2002}. 

In literature, few studies are present which are using multiple IHC for analysis and also using them to examine H\&E such as \textit{Bulten et al.}~\cite{Bulten2019} and \textit{Solorzano et al.}~\cite{Solorzano2018}. They are using color deconvolution technique to separate Hematoxylin and Diaminobenzidine (H and DAB), which is not very accurate when it comes to the diagnosis of HL. On the other hand, technique such as multiplex IHC is used to examine co-localisation of expression of markers~\cite{Denise2014}. However, multiplexed IHC is a complex process and are available only in advanced laboratories~\cite{VanHerck2021}.

In this study, we performed preliminary experiment on HL by proposing composite biomarker image (CBI) technique, which is composed by superimposing multiple IHC biomarker WSIs. The purpose of this CBI is to assist pathologists with diagnosis accuracy and speed up the process. Furthermore, the pathologists can visualize the selected information from multiple biomarker images in one single image and they don't have to go back and forth to analyze different images, thus avoiding any confused diagnosis.

The rest of the paper is organized as follows: In Section~\ref{sec:method},
the proposed method is presented along with the data that we used. In Section III, the results from the proposed method are presented. Section IV includes the conclusion and future work insight.

\section{MATERIALS \& METHODS}
\label{sec:method}
This paper presents a novel CBI image produced in two steps by the proposed system. In the first step, the source biomarker images are registered according to the target H\&E image. Furthermore, in the second step, the specific regions (based on the pathologists' recommendation) from each biomarkers are filtered using fuzzy inference system (FIS), and combined into one image. Fig.~\ref{Fig:Methods} illustrates the overall two step process followed by the proposed system to generate a CBI image and attention map.

\textbf{Adaptive neuro-fuzzy inference system (ANFIS) -- }
%\label{Sec:Fuzzy_System}
First of all, biomarker images are aligned together according to a single coordinate system so that the tissue could be aligned~\cite{Abubakr2021}. After aligning the biomarker images, FIS is used to filter and extract the positive or negative information from the biomarker images at single cell level. Here, we used Adaptive neuro-fuzzy inference system (ANFIS) to mimic the pathologists observation about each biomarker, about what they specifically look for in each WSI. ANFIS is the combination of neural networks (NN) and FIS which uses NNs learning algorithms to self tune the rules, parameters, and structure of FIS~\cite{Alizadeh2017, Hosseini2012}. For simplicity, we assume a network with three inputs (channels of an image) $r$, $g$ and $b$, and one output, $o$~\cite{jang1992anfis, Hosseini2012}. To present the ANFIS architecture, three fuzzy if–then rules based on a first-order Sugeno model are shown below:

\noindent $\textbf{Rule 1:}$ $if$ $r$ $is$ $A_1$ $and$ $g$ $is$ $B_1$ $and$ $b$ $is$ $C_1,$ $then$ $o_1$ $=j_1 r + k_1 g + l_1 b + z_1$\\
\noindent $\textbf{Rule 2:}$ $if$ $r$ $is$ $A_2$ $and$ $g$ $is$ $B_2$ $and$ $b$ $is$ $C_2,$ $then$ $o_2$ $=j_2 r + k_2 g + l_2 b + z_2$\\
\noindent $\textbf{Rule 3:}$ $if$ $r$ $is$ $A_3$ $and$ $g$ $is$ $B_3$ $and$ $b$ $is$ $C_3,$ $then$ $o_3$ $=j_3 r + k_3 g + l_3 b + z_3$

\noindent where $r$, $g$, and $b$ are the inputs, $A_i$, $B_i$, and $C_i$ are the fuzzy sets, $o$ is the output of the fuzzy system, and $j_i$, $k_i$, $l_i$, and $z_i$ are the design parameters that are calculated during the training process.

\begin{figure*}[ht]
\centerline{\includegraphics[width =0.90\textwidth]{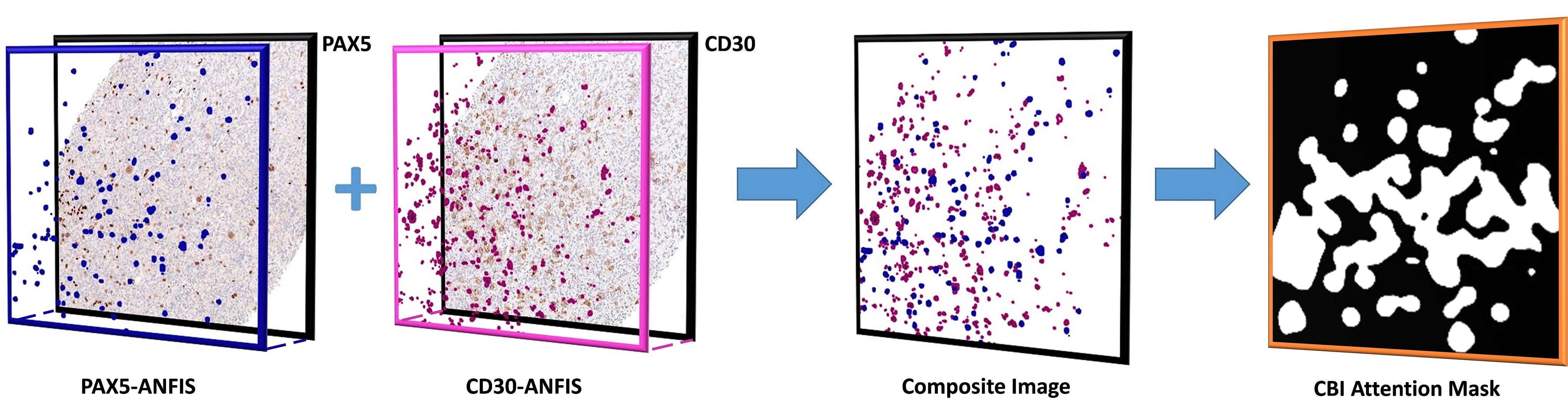}}
\caption{Sample CD30 positive cell population, and PAX5 semi-positive cell population filtered from the registered area using ANFIS. Filtered ANFIS images are them combined to generate a CBI image and eventually a CBI Mask (attention map) is generated using common tissue regions.}
\label{Fig:CBI}
\end{figure*}

In our experiment, FIS are used as a filter for the input IHCs (see Fig.~\ref{Fig:Methods} \& Fig.~\ref{Fig:CBI}). FIS filter the WSI at pixel level based on thee RGB color values, thus taking three inputs as R, G, and B and giving one output $o$ in the range of $0$ to $255$ as a gray scale image. The FIS uses biomarker specific models, which are trained using a range of RGB data collected randomly by our experts (see table~\ref{tab:Class}). For instance, in CD30, only positive regions (class 3, 4, 5 from table~\ref{tab:Class}) are to be extracted, and in PAX5 semi-positive tissues (class 4 only from table~\ref{tab:Class}) are to be retrieved, hence only the target tissues will get some dark grey levels and rest of the area will be assigned as background (white color). In the end, a gray image is generated as an output ($o$) containing only the gray impressions of target tissues (see fig~\ref{Fig:CBI}).

\begin{table}[htbp]
  \centering
  \caption{FIS classification training input data range of \textbf{R, G, and B} pixels and output classes.}
\begin{tabular}{|c|c|c|l|}
\hline
\multicolumn{3}{|c|}{\begin{tabular}[c]{@{}c@{}}\textbf{Input Range}\end{tabular}} & \multicolumn{1}{c|}{\multirow{2}{*}{\textbf{Class}}} \\ \cline{1-3}
\textbf{R}                        & \textbf{G}                        & \textbf{B}                       & \multicolumn{1}{c|}{}                       \\ \hline
214 - 247                & 214 - 247                & 213 - 247               & 0 = Background                              \\ \hline
24 - 208                 & 44 - 217                 & 79 - 228                & 1 = Blue                                    \\ \hline
63 - 221                 & 65 - 221                 & 77 - 226                & 2 = Gray                                    \\ \hline
163 - 251                & 124 - 218                & 107 - 214               & 3 = Light Brown                             \\ \hline
136 - 192                & 87 - 157                 & 70 - 147                & 4 = Medium Brown                            \\ \hline
37 - 98                  & 0 - 67                   & 0 - 61                  & 5 = Dark Brown                              \\ \hline
\end{tabular}
\label{tab:Class}%
\end{table}

\textbf{Composite Biomarker Image --}
Final Step of the proposed method, after FISs, is to generate a CBI image. After getting the gray level images of each biomarker with the tissues of interest, these images are morphologically processed and merged together to form a CBI image. Filtered biomarker images are arranged and overlayed on each other. In our experiment, we arranged CD30 as the first image, then PAX5 on top of CD30. A sample CBI image is shown in Fig.~\ref{Fig:CBI} for illustration. The overlay color of the biomarker images can be chosen as per convenience, and their overlay order in CBI image can also be adjusted according to the needs. Furthermore, an attention map (mask) can also be generated from the CBI image based on the common population density of the filtered tissues among multiple biomarkers (see Fig.~\ref{Fig:CBI}).

\textbf{Data -- }
In the experiment, 12 cases (or 36 WSIs) of CHL are used, along with H\&E and the relevant IHC biomarkers (CD30 and PAX5). The images were acquired from the Grand River Hospital, Kitchener, Ontario, Canada using Huron Tissue Scope LE scanner. The specimen are taken from different parts of the body. These specimens should be cut into small parts and each part will be fit to a paraffin cassette. From each cassette, H\&E and the IHCs are prepared on a glass slide, which are further scanned at 40x magnification to obtain the digital WSI. 

For the ANFIS training and testing data, 30 RGB values were randomly selected by the expert from the area of each class, so in total 180 x 3 (RGB) points were used in our experiment. From 30 points, 25 RGB points were used for the training, and 5 points were used for testing for each class. The range of RGB data for each class for training is shown in Table.~\ref{tab:Class}.

\begin{figure}[ht]
\centerline{\includegraphics[width = 0.5\textwidth]{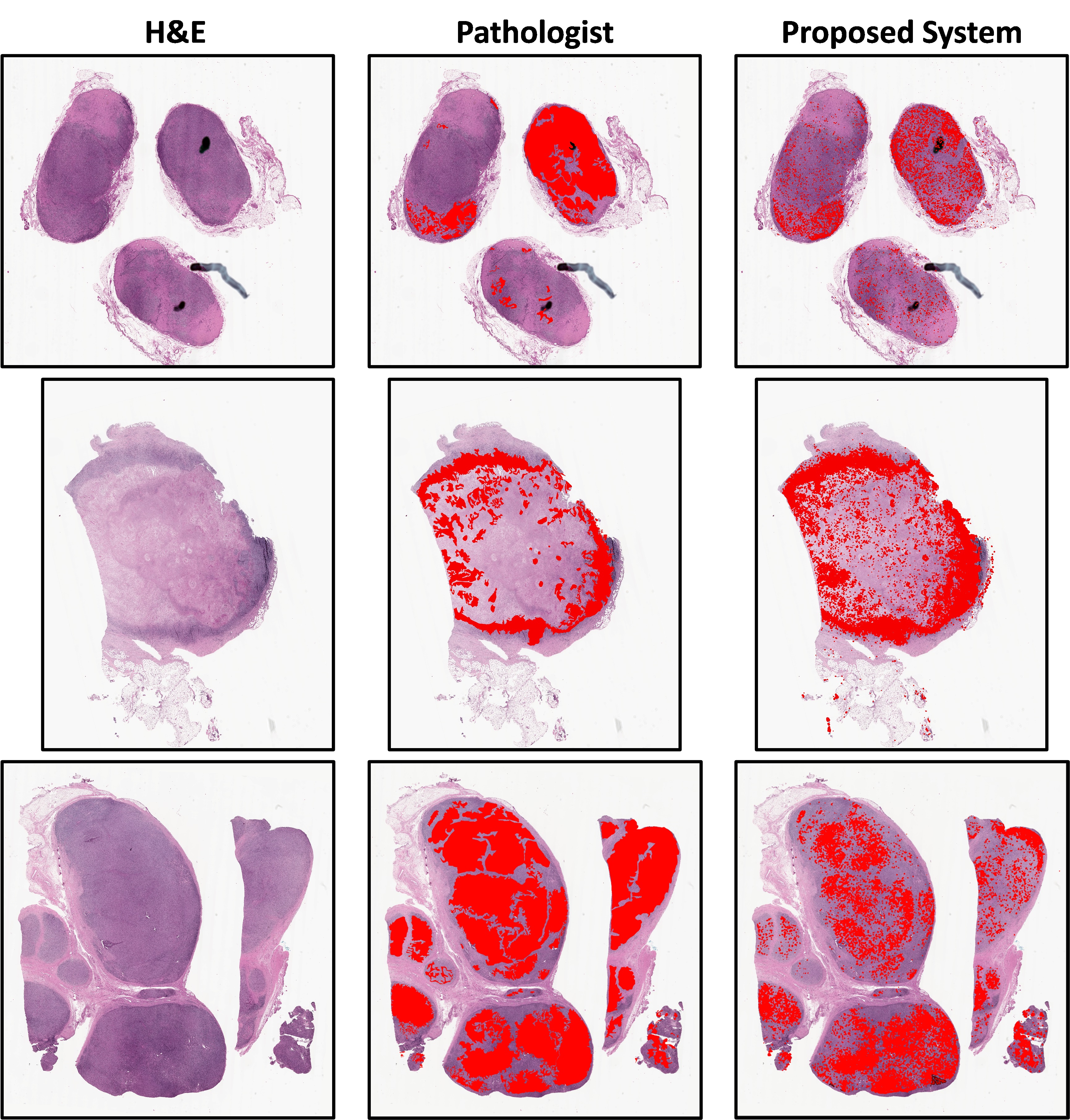}}
\caption{A qualitative comparison between the expert's annotated masks and the attention mask generated by the proposed CBI system.}
\label{Fig:Result3}
\end{figure}

\section{Results}
\label{Sec:rslt}

\begin{figure*}[ht]
\centerline{\includegraphics[width =0.89\textwidth]{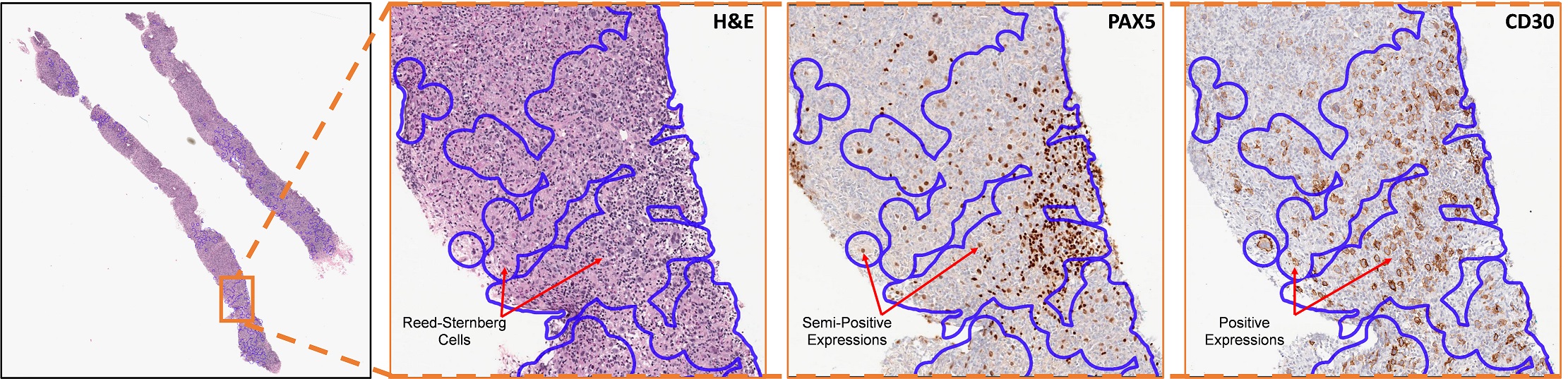}}
\caption{A sample H\&E WSI overlayed by the automated CBI attention mask from the two biomarker images. From the common annotated region, we can identify the Reed-Sternberg cells, Semi-positive expressions, and positive expressions in H\&E, PAX5, and CD30 images respectively.}
\label{Fig:Results}
\end{figure*}

The proposed system was evaluated in this preliminary study and diagnosis of CHL. 12 CHL cases were selected and examined by the expert pathologists from the Grand River Hospital, Kitchener, Canada and the Mayo Clinic, Rochester, USA. The purpose of CBI is to aid the pathologists with their diagnosis with the help of multiple biomarkers and selected regions. Fig.~\ref{Fig:CBI} shows a sample CBI image produced by the proposed system using two IHCs which include CD30 positive, and PAX5 semi-positive. The number of biomarkers can vary according to the diagnosis requirements. However, in our experiment, we have used two biomarkers (CD30, and PAX5) for CHL. Moreover, our proposed system takes full advantage of registered IHCs and take the common region where the CD30 have positive tissue population and PAX5 have semi-positive tissue population to generate an attention map which shows the possible regions containing the CHL malignant cells (see Fig.~\ref{Fig:Results}). Furthermore, the results from the proposed system were analysed qualitatively by the expert pathologists, and the map generated by the proposed system seems to be more accurate and precise at cellular level when compared to the pathologists marked regions (as shown in Fig.~\ref{Fig:Result3}). In this comparison, pathologists analyse the biomarkers (CD30, and PAX5) and annotated according to their findings. Pathologist's delineation may include some human error, and some benign/healthy tissue. For our experiments, we were unable to generate a similarity indexes because the ground truth labelled by the pathologists and the maps generated by the proposed system were very different in terms of precision, but they majorly covered the same area as we can observe in Fig.~\ref{Fig:Result3}. Therefore, in this case, the similarity index could be misleading. Thus, qualitatively evaluating the CHL maps from the proposed system.

\section{Conclusions}
\label{Sec:conclusion}
In this paper, we introduced CBI as an advanced visualization modality, which can assist the pathologists to identify the key suspected areas, which otherwise would be a cumbersome task involving  inspection of multiple IHC WSIs back and forth from one biomarker image to another. Moreover, IHC images are typically not aligned with each other, so pathologists also have to identify the co-localized areas visually. The proposed CBI enables the experts to read the entire information from a unified modality.  For the future work, one should exploit CBI in applications where extensive annotations from the experts are required. Additionally, we need figure out how to quantitatively compare the results of the proposed system with those of the pathologists.

% References
\bibliographystyle{IEEEtran}
\bibliography{CBI_Ref.bib}

\end{document}